\documentclass{elsart}

\usepackage{natbib}

\usepackage{graphicx}

\usepackage{amssymb}

\begin{document}

\begin{frontmatter}



\title{A Monte Carlo approach to study
neutron and fragment emission in heavy-ion reactions}


\author[miuni,minfn]{M.V. Garzelli\corauthref{cor1}}
\corauth[cor1]{Corresponding author.}
\ead{garzelli@mi.infn.it}
\author[minfn]{P.R. Sala}
\author[pvunin]{F. Ballarini}
\author[minfn]{G. Battistoni}
\author[cern]{F. Cerutti}
\author[cern]{A. Ferrari}
\author[miuni,minfn]{E. Gadioli}
\author[pvunin]{A. Ottolenghi}
\author[hou]{L.S. Pinsky}
\author[sie]{J. Ranft}

\address[miuni]{University of Milano, Department of Physics, 
via Celoria 16, I-20133 Milano, Italy}
\address[minfn]{INFN, Sezione di Milano, via Celoria 16, I-20133 Milano, Italy}
\address[pvunin]{University of Pavia, Department of Theoretical and Nuclear Physics, and INFN, Sezione di Pavia, via Bassi 6, I-27100 Pavia, Italy}
\address[cern]{CERN, CH-1211 Geneva, Switzerland}
\address[hou]{University of Houston, Department of Physics, 
617 Science \& Research Bldg 1, Houston, TX 77204-5005, US}
\address[sie]{Siegen University, Fachbereich 7 - Physik, 
D-57068 Siegen, Germany}
\begin{abstract}
Quantum Molecular Dynamics models (QMD) 
are Monte Carlo approaches targeted at the
description of nucleon-ion and ion-ion collisions. We have developed
a QMD code, which has been used for the simulation of the fast stage
of ion-ion collisions, considering a wide range of system masses and
system mass asymmetries. The slow stage of the collisions has been
described by statistical methods. The combination of both stages
leads to final distributions of particles and fragments, which have
been compared to experimental data available in literature. 
A few results of these comparisons, 
concerning neutron double-differential production cross-sections
for C, Ne and Ar ions impinging on C, Cu and Pb targets 
at 290 - 400 MeV/A bombarding energies  
and fragment isotopic distributions 
from Xe + Al at 790 MeV/A, are shown in this paper.
\end{abstract}

\begin{keyword}
heavy-ion collisions \sep QMD \sep fragmentation \sep evaporation
\PACS 24.10-i \sep 25.70.Mn \sep 21.60.Ka
\end{keyword}

\end{frontmatter}

\section{Introduction}
\label{intro}
Since several years, Quantum Molecular Dynamics models (QMD) 
have been introduced 
to simulate heavy-ion 
collisions (see e.g. \citet{aiche} and re\-fe\-ren\-ces therein, 
 \citet{jap}). 
They are semi-classical approaches (\citet{cana}),
derived from the Classical Molecular Dynamics ones, by adding a few quantum
features. The main one is that each nucleon is described by a 
coherent state, and is thus identified by gaussian spatial coordinate
and momentum distributions, centered around mean values, and characterized
by gaussian widths whose product minimizes the Heisenberg 
uncertainty relationship. A second quantum feature 
is the fact that nucleon-nucleon scattering cross-sections taken 
from the experiments are used to 
describe nucleon-nucleon collisions occurring in the fast-stage 
of the reactions between ions, when nuclei can overlap, depending on the
impact parameter value. 
The impact parameters are randomly chosen,
according to a Monte Carlo procedure. Thus both very peripheral reactions
and more central ones can occur, on an event by event basis.  
Actually the measured free nucleon-nucleon cross-sections are used when
a nucleon-nucleon collision occurs in these reactions, while   
the effects on each nucleon of all other nucleons 
present both in the projectile ion and
in the target one are taken into account 
by means of effective potentials, included in the interaction part 
of the Hamiltonian, during the whole simulation.  
Another quantum feature is the fact that the collisions which lead a nucleon
in an already occupied phase-space region are forbidden, 
according to the
Pauli blocking principle. This is especially effective at non-relativistic
e\-ner\-gies, where nucleon-nucleon potentials are increasingly
more important than nucleon-nucleon collisions
to determine the fragmentation process, since most 
nucleon-nucleon collisions are blocked.  
Finally, a few authors have also de\-ve\-lo\-ped fully antisymmetrized versions
of QMD models (e.g. Fermionic Molecular Dynamics by~\citet{fmd1, fmd2}, 
and Antisymmetrized Molecular 
Dynamics by~\citet{amd1, amd2}), 
by fully antisymmetrizing the nuclear wave-functions, which
in the original QMD versions are merely given by the product of nucleon
wave-functions. At present, these advanced versions are succesfully used 
in very specific problems, such as the study of exotic nuclear structure, 
mainly for light nuclei
due to the complexity of the underlying approaches, 
and their intensive CPU  requirement. 
We limit to the original approximation for the nuclear
wave-functions, instead of using a more advanced approach, since
our final aim is developing a code to be used real-time 
to microscopically describe ion-ion collisions
in general applicative problems, which can involve even
the heaviest ions and composite targets. To accomplish this aim,
a preliminary version of our code has been interfaced to the general 
purpose Monte Carlo FLUKA code, allowing the simulation of fragment 
formation and the study of the emission of fragmentation products,  
in thin and thick targets, even in complex geometries.
Nevertheless, to better characterize and investigate
the nuclear reaction processes,
the results of si\-mu\-la\-tions 
performed in the simplest theoretical case of generating 
thousands of single ion-ion collision events
 are shown in this paper. This means that ion-ion 
overlapping and nuclear de-excitation
effects have been accurately considered, while target thickness effects, 
such as ion energy loss and reinteraction in matter,
were neglected in making these simulations. 

\section{Theoretical elements of our simulations}
\label{theo}
A QMD code has recently been  developed by our Collaboration
and interfaced to the FLUKA code. 
FLUKA (\citet{flukacern}) is a particle tran\-sport and interaction code, 
used since several years
in many fields of Physics. It is currently developed 
according to an  INFN - CERN official agreement
and is made available on the web ({\it http://www.fluka.org}). 
It is based, as far as possible, on 
theoretical microscopic models, which are continuously updated and
improved, and benchmarked with respect to newly available experimental
data. Nucleon-nucleus collisions are simulated in FLUKA by means of the 
module called PEANUT (PreEquilibrium 
And NUclear Thermalization,~\citet{trieste}),
which contains also a de-excitation sub-module. This sub-module has also
been used to describe the de-excitation processes which may occur
at the end of the overlapping stage of ion-ion collisions, which
has been simulated in this work by means of our QMD code.
In particular, the QMD code has been used 
to follow the ion-ion system evolution
for a few hundreds of fm/c, while de-excitation processes can occur
on a time scale far larger (up to $\sim 10^7$ fm/c), and have been 
simulated by means of the statistical models contained 
in the FLUKA de-excitation sub-module. De-excitation is provided by different
mechanisms, sometime in competition. Evaporation has been considered,
both of light particles and fragments, up to a mass~A~=~24. 
Fission processes 
can occur, as well as Fermi break-up of light nuclei, with a maximum
of six bodies in the final state in the present implementation. 
Finally, de-excitation by photon emission
is also included in FLUKA.  
A correct description of de-excitation is crucial, since it deeply
modifies the particle and fragment 
pattern present at the end of the fast stage 
of ion-ion collisions. In fact, while peripheral collisions give often
rise to slightly excited projectile-like and target-like fragments, 
with masses very similar to the
projectile and target ones, respectively,
very central collisions are capable of generating 
intermediate mass
fragments (IMF), characterized by lower masses and higher excitation
energies: hot fragments in turn decay, deeply affecting the final
fragment and particle emission distributions. 
A few features of our QMD code are presented in Subsection~\ref{qmd},
while the results of our simulations are shown in Section~\ref{results}.

\subsection{QMD specific features}
\label{qmd}
As explained in the Introduction, several 
QMD versions and codes have already been developed,
characterized by some common features, and some specific differences,
depending on the working group and the intended application. 
Here some features of the code we have developed are briefly sketched.
The Hamiltonian is made by a kinetic and an effective interaction part.
Our code can switch between non-relativistic and relativistic kinematics,
according to a user initial choice, while only instantaneous interactions are
included. A fully relativistic version, including relativistic potentials,
is also under development thanks to the cooperation with the University of
Houston and NASA, and is the object of a separate work (\citet{zapp}).
Neutrons and protons are distinguished both for their
interaction and for their mass. The initial nuclear states, used in
the ion-ion collision simulation processes, are prepared in advance, stored
off-line and exactly fullfill the experimental constraints
on bin\-ding energies. 
Binding energies are calculated by summing the kinetic energy 
and the potential energy of all nucleons in the nucleus. The potential
energy is given by a superposition 
of different terms, due to the interaction of
each nucleon with all other nucleons present in the ion. In particular,
Skyrme-type two-body and three-body interaction terms are included,
as well as two-body symmetry and surface terms. The symmetry term helps
taking into account the difference among neutrons and protons, by assigning
different weights to the interactions between nucleons with the same isospin 
and nucleons with different isospin. Furthermore, a Coulomb term 
is included, which acts only on protons. Further details on the effect
of each term in determining the total potential energy of each ion can
be found in~\citet{garzvar06}. When two ions collide, each nucleon is 
subject to the effects of both the nucleons in the same nucleus and those
in the other one. At ion-ion distances ex\-cee\-ding a few fm, 
only the Coulomb force is important,
while nuclear interaction among nucleons of different nuclei
becomes significant only at lower distances, i.e. during the ion-ion 
overlapping process.
Also nucleon-nucleon collisions can occur in this stage, which are
implemented in our code taking into account free  {\it n}-{\it p} and
{\it p}-{\it p} ({\it n}-{\it n}) reaction
cross-sections. As an approximation, this process is assumed 
to be isotropic, i.e. the  
angular dependendence of the cross-sections is neglected, as well as
the pion production channel. Further extensions of the model will include
these elements.  
Ion-ion system evolution is followed for
a time of $\sim 150 - 200$ fm/c after the overlapping stage.
At the end, 
hot excited fragments may be present, as well as emitted
neutrons and protons. The de-excitation of hot fragments has been simulated
thanks to the interface to the FLUKA de-excitation module, as explained
at the beginning of Section~\ref{theo}.     
 
\section{Results of the simulations and comparisons with experimental data}
\label{results}
As explained in Section~\ref{theo}, 
the results presented in this paper have been obtained by the interface
between our QMD code and the FLUKA nuclear de-excitation routines. 
In Subsection~\ref{neut}
double-differential neutron production cross-sections for several
projectile-target systems at energies 290 and 400 MeV/A are compared
to available experimental data, while in Subsection~\ref{isot} fragment
isotopic distributions from Xe projectiles impinging on an Aluminium target
at 790 MeV/A are shown. QMD + FLUKA has also been tested at energies
around and below 100 MeV/A. A few results of these lower-energy tests
have been presented in~\citet{paviamv, garzvar06, garzrila06}.   

\subsection{Double-differential neutron production cross-sections}
\label{neut}
\begin{figure}[p!]
\begin{center}
\includegraphics[bb=51 51 613 793,  width=0.32\textwidth]{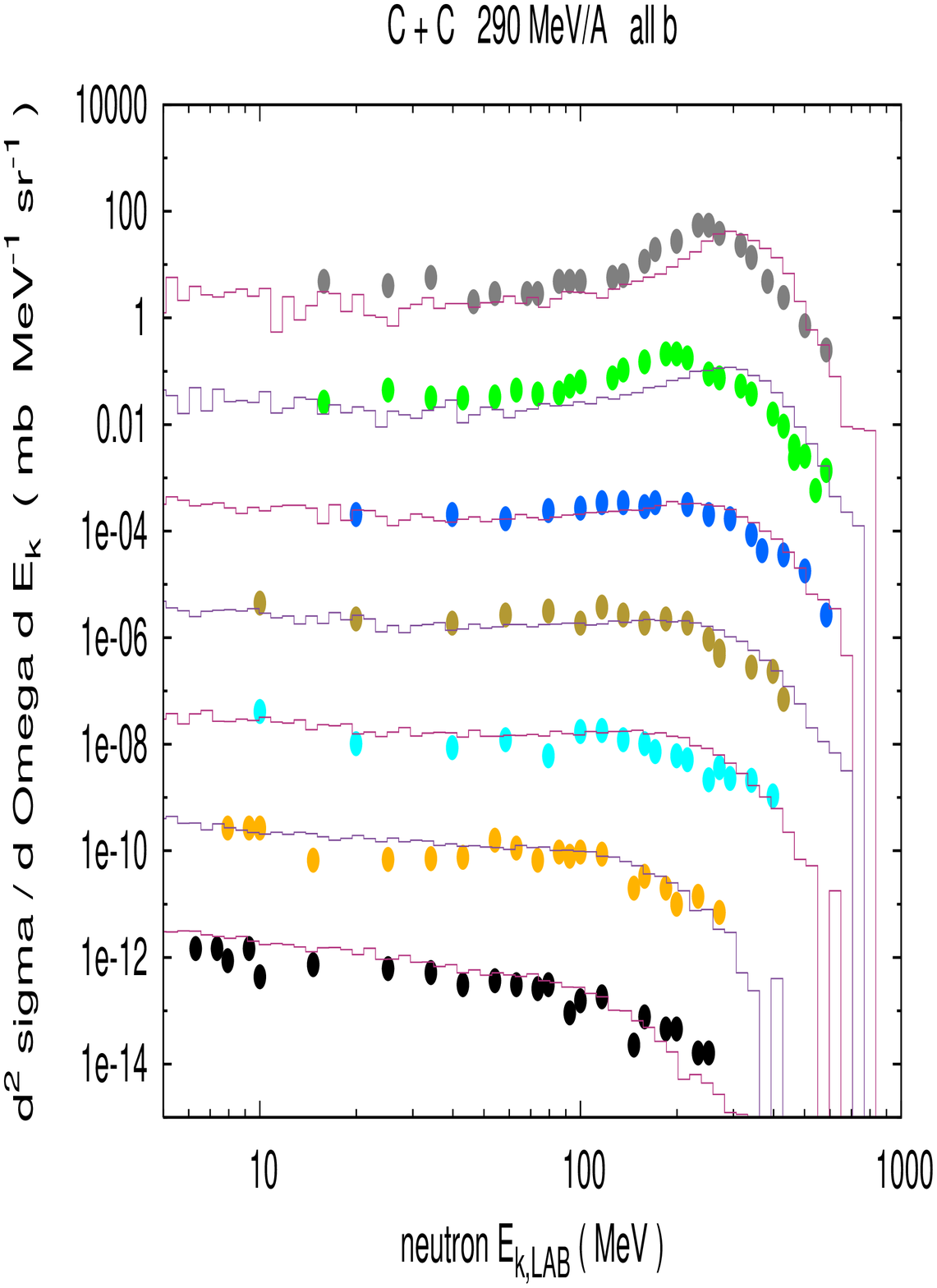}
\includegraphics[bb=51 51 613 793,  width=0.32\textwidth]{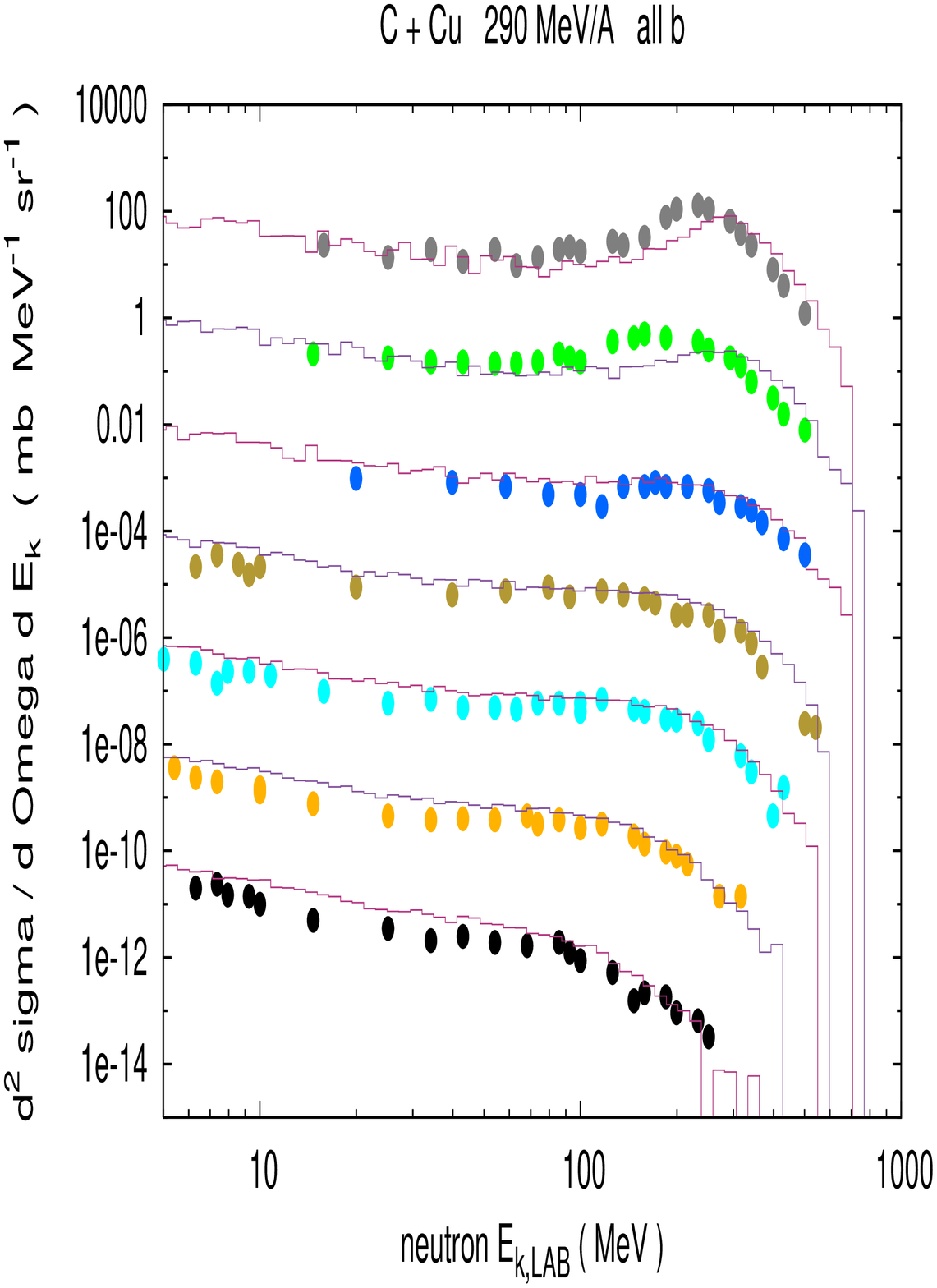}
\includegraphics[bb=51 51 613 793,  width=0.32\textwidth]{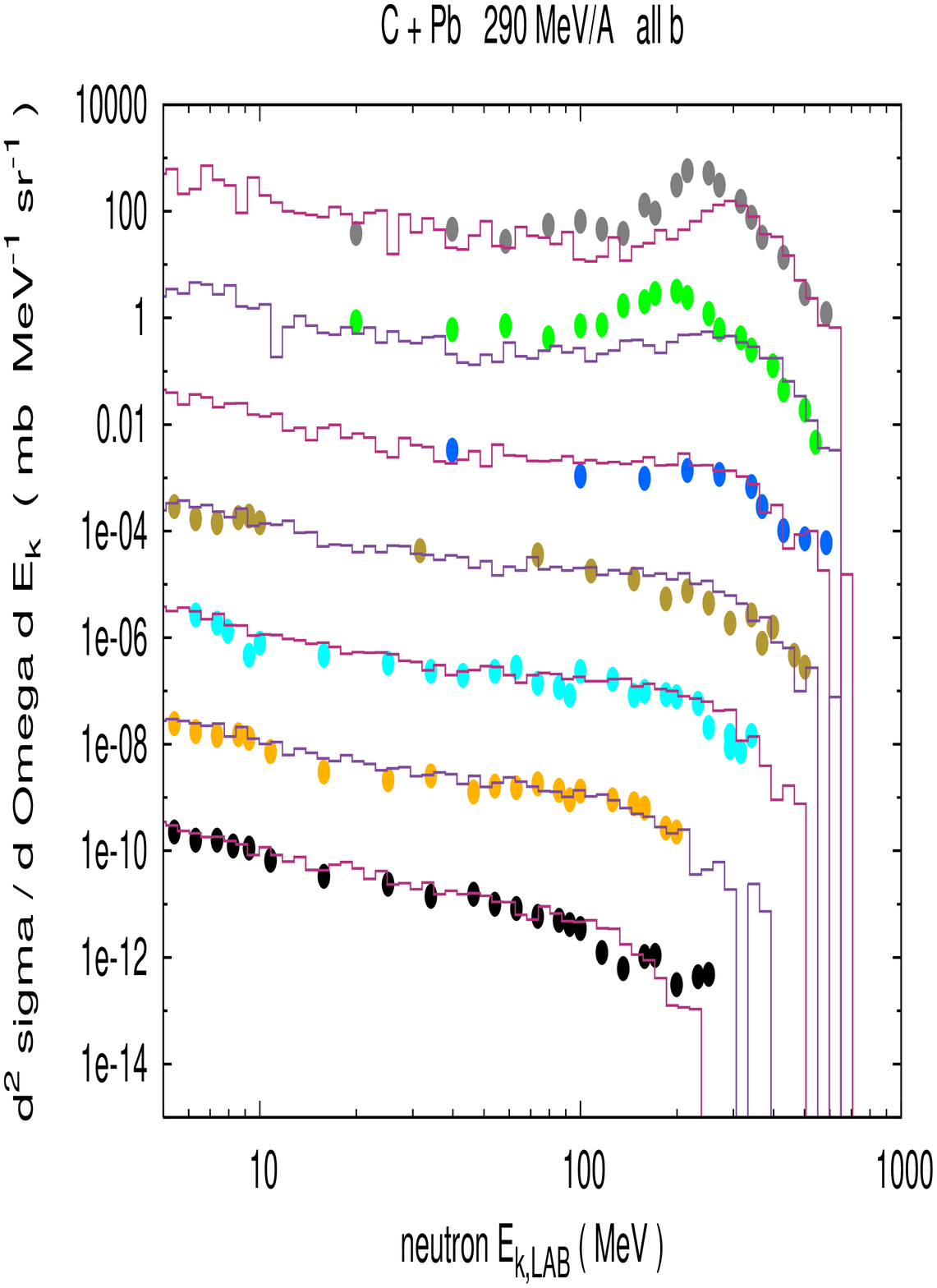}
\\
\includegraphics[bb=51 51 613 793,  width=0.32\textwidth]{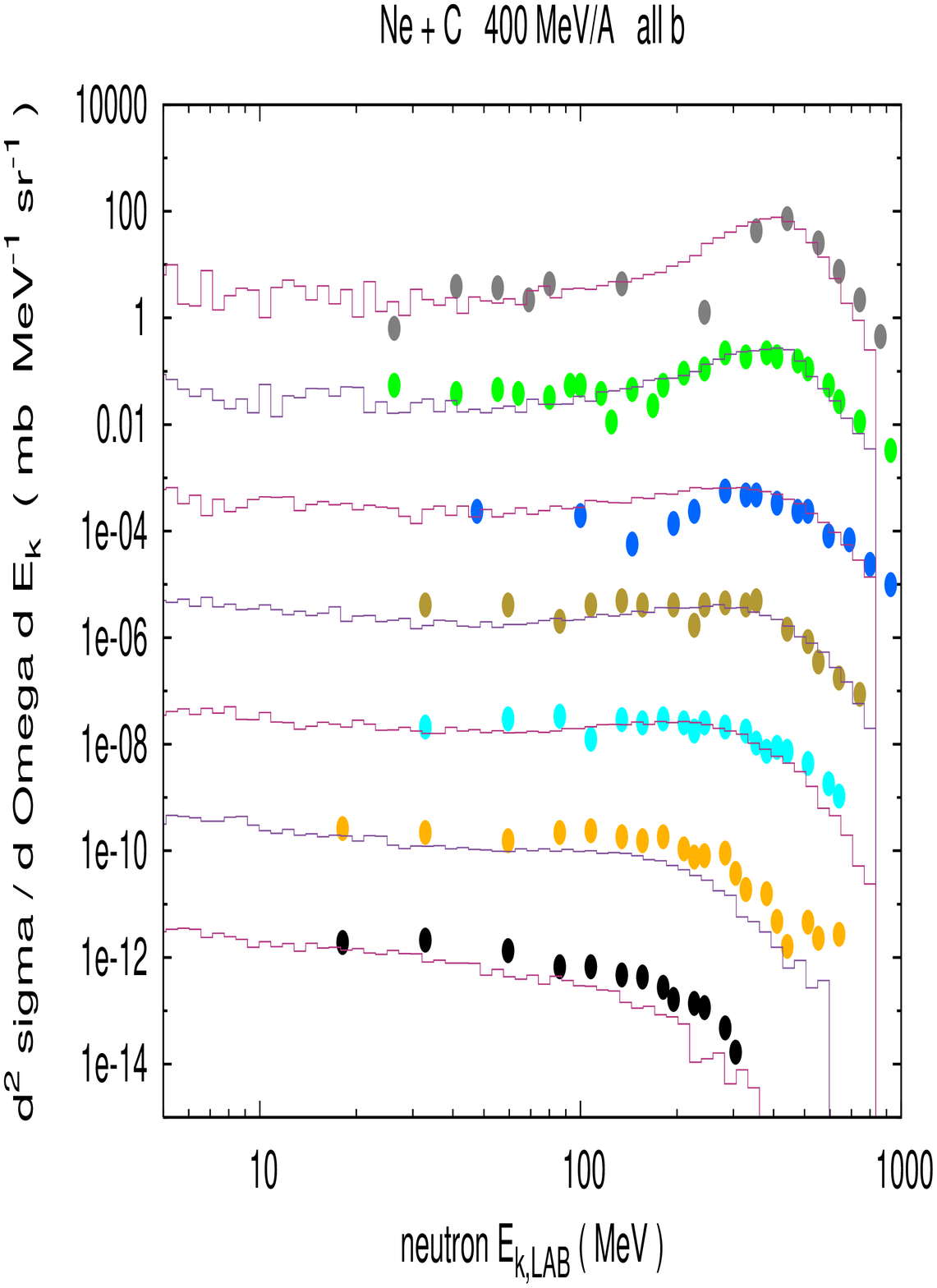}
\includegraphics[bb=51 51 613 793,  width=0.32\textwidth]{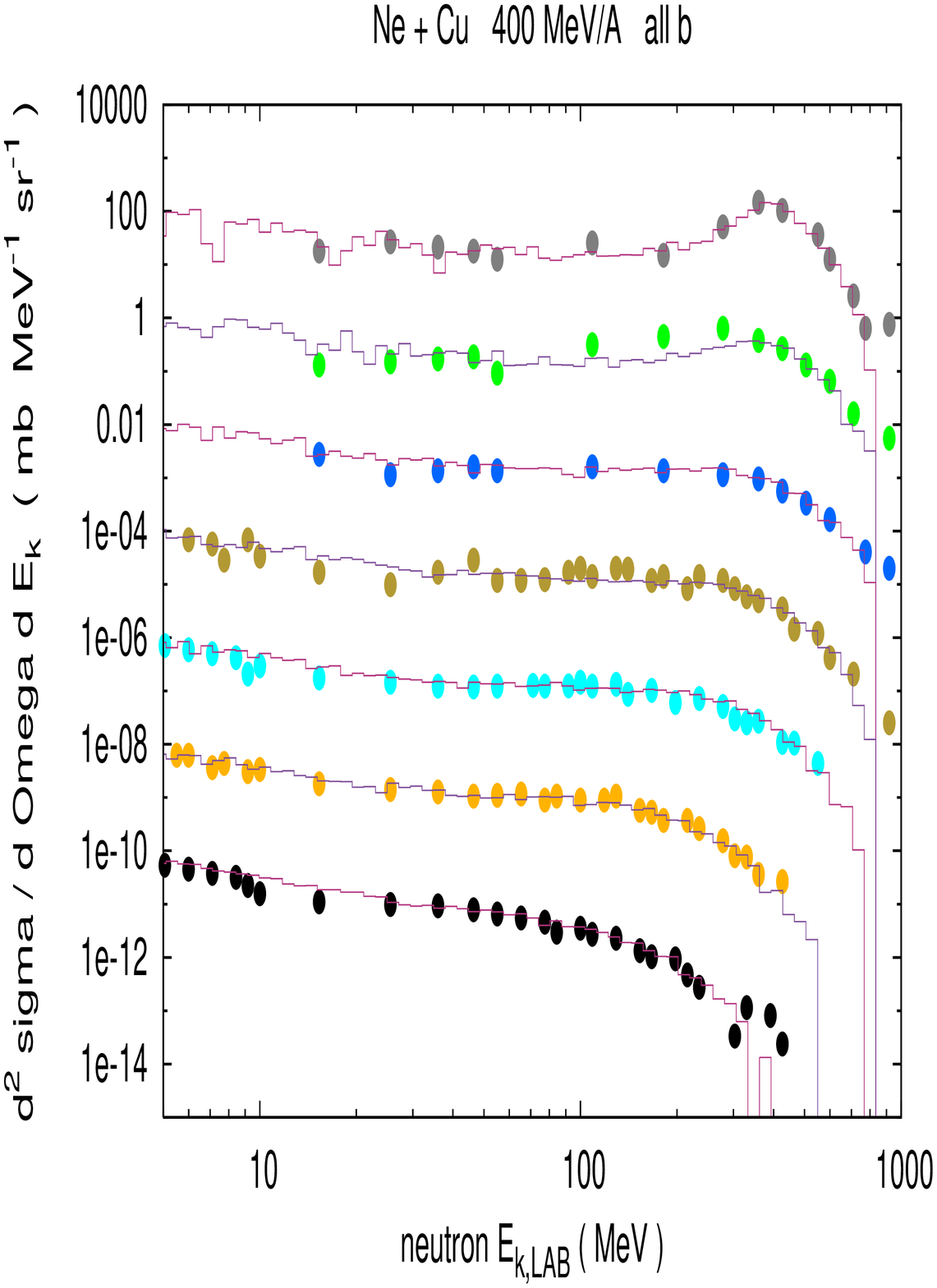}
\includegraphics[bb=51 51 613 793,  width=0.32\textwidth]{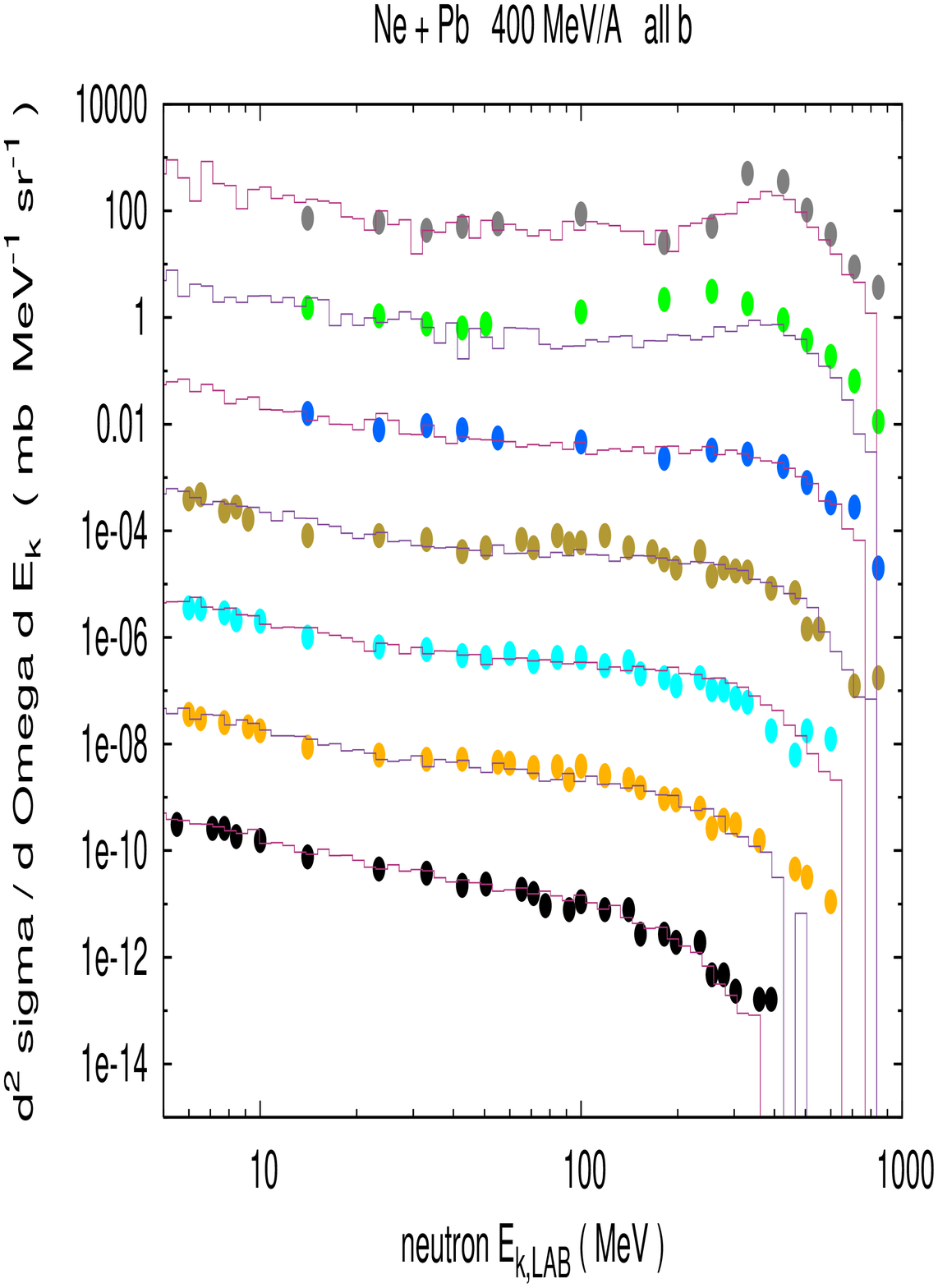}
\\
\includegraphics[bb=51 51 613 793,  width=0.32\textwidth]{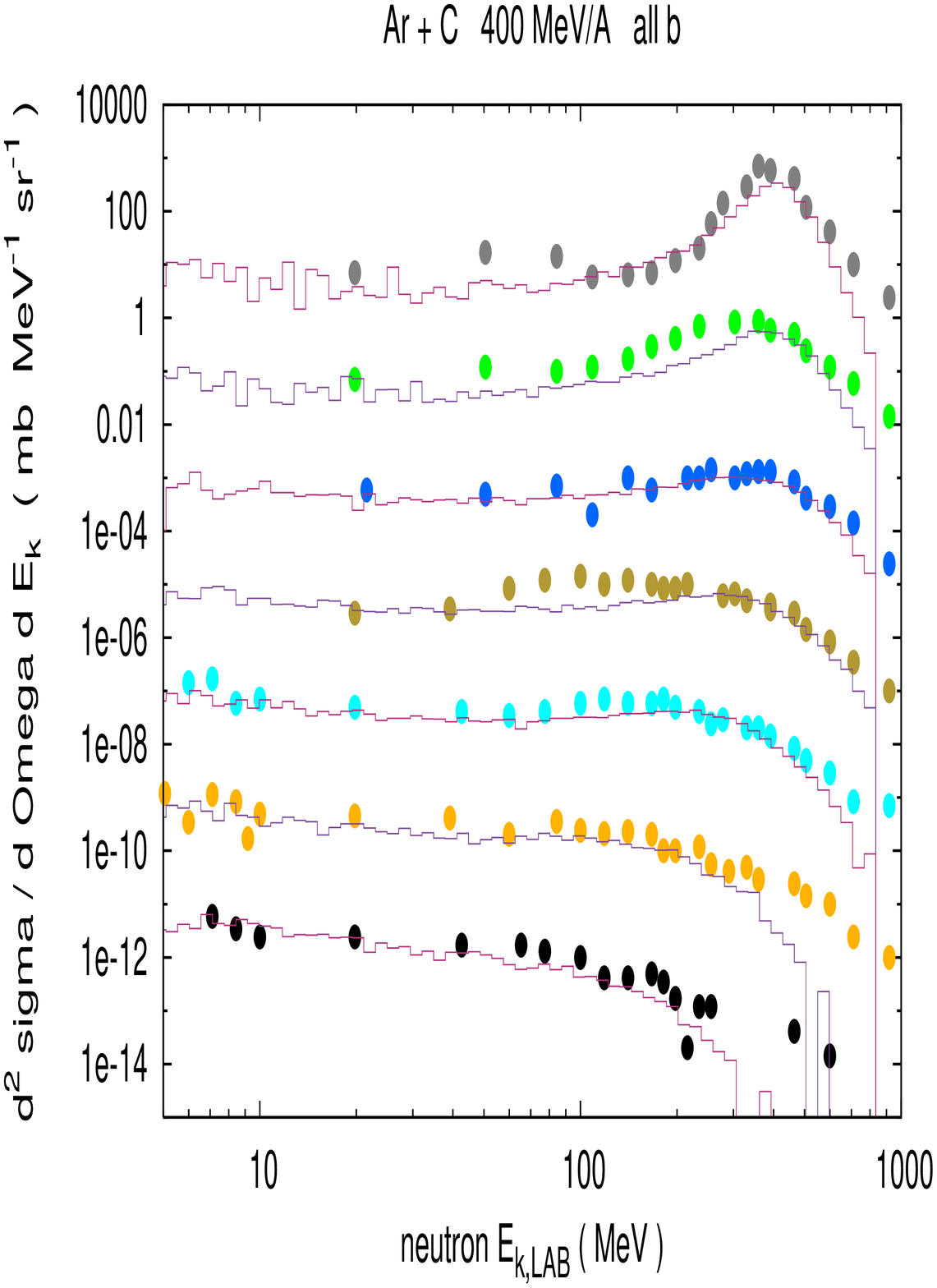}
\includegraphics[bb=51 51 613 793,  width=0.32\textwidth]{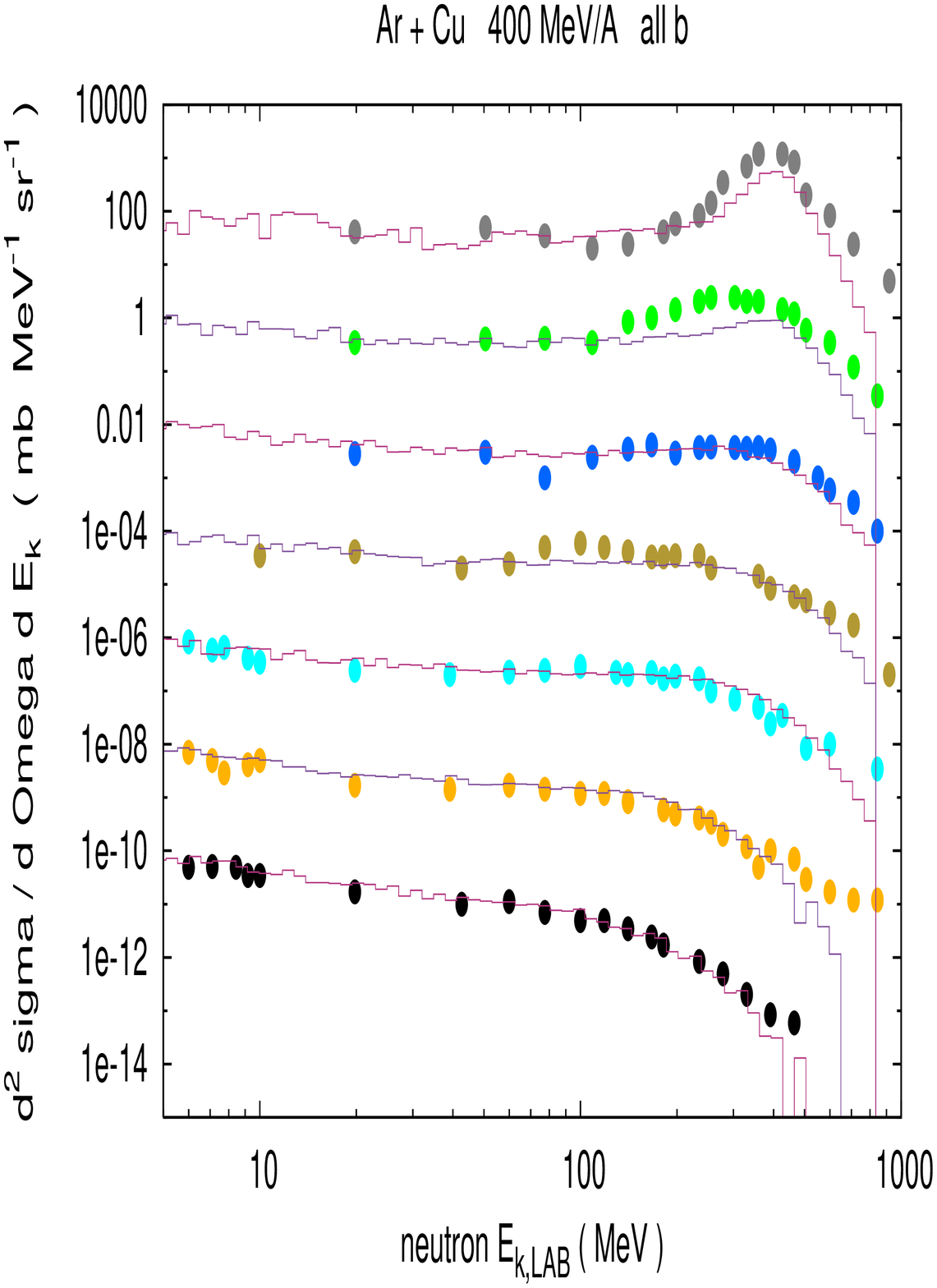}
\includegraphics[bb=51 51 613 793,  width=0.32\textwidth]{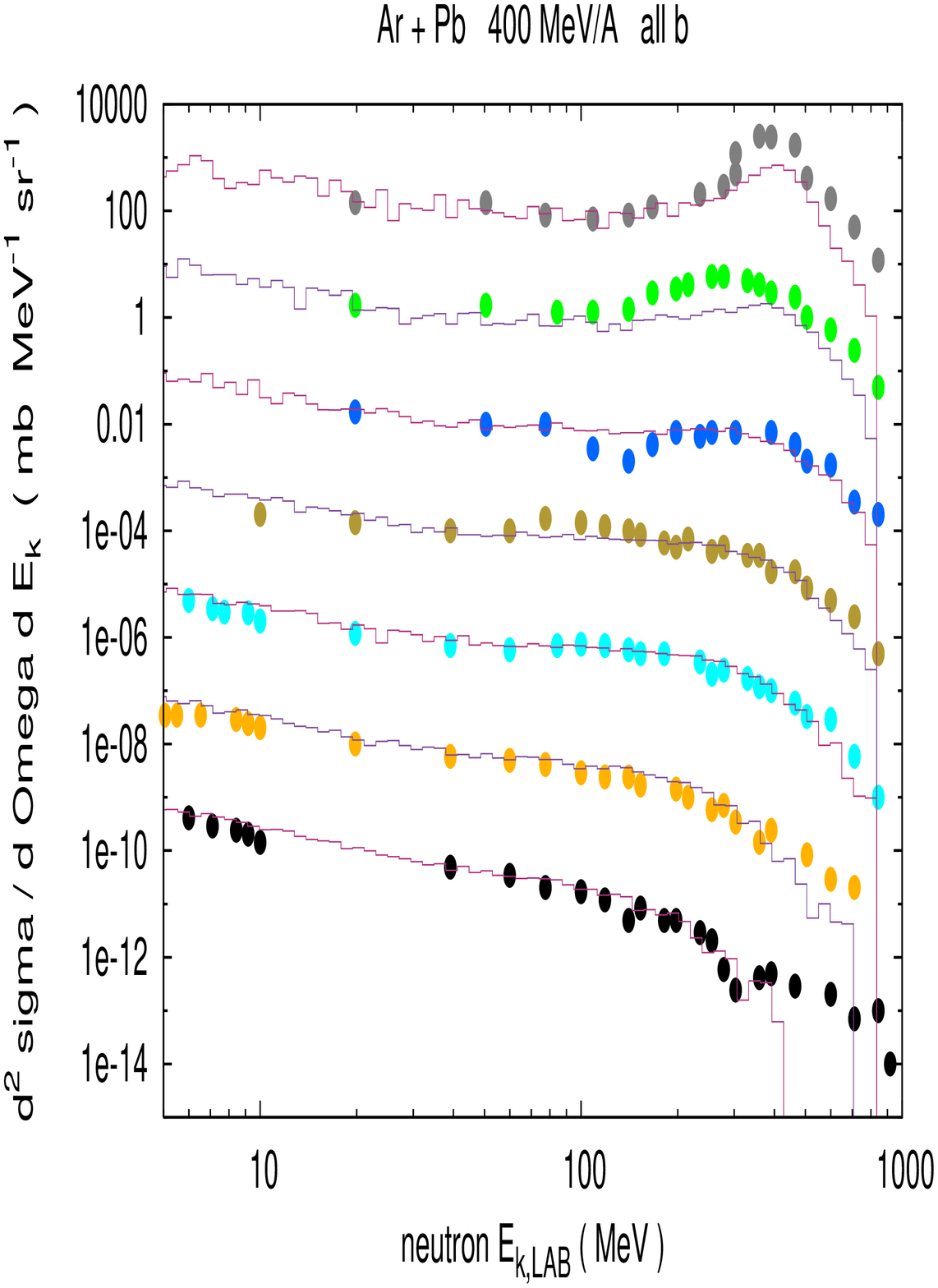}
\end{center}
\caption{\label{figneu} Double-differential neutron production cross-section
for C (top), Ne (center) and Ar (bottom) projectiles impinging on
C (left), Cu (center) and Pb (right) 
at 290, 400, 400 MeV/A bombarding energies, respectively. 
The results of the theoretical simulations made by QMD + FLUKA are shown
by solid histograms, while the experimental data measured by~\citet{iwata}
are shown by filled circles. In each panel, distributions at
5$^{\mathrm{o}}$, 10$^{\mathrm{o}}$, 20$^{\mathrm{o}}$, 
30$^{\mathrm{o}}$, 40$^{\mathrm{o}}$, 60$^{\mathrm{o}}$ and 
80$^{\mathrm{o}}$ angles with respect to the incoming beam direction
in the laboratory frame, 
have been multiplied by decreasing even powers of 10, 
for display purposes. The details 
of target geometry (thickness and shape) have not been taken into 
account in our theoretical simulations.}
\end{figure}

Several papers presenting neutron double-differential production
cross-section data have been published. Most of them consider thick targets.
Nevertheless, at the aim of investigating the reliability and 
the prediction capabilities 
of theoretical ion-ion reaction models, 
it is better disentangling transport effects, 
thus data from thin target experiments are more suited. 
~\citet{iwata} present double-differential neutron production
cross-sections for
many systems, mostly considering thin targets. 
Therefore, the experimental
data presented by these authors have been considered for our simulations.
The targets used in their experiment
are made of Carbon, Copper and Lead. The results of our simulations
for a Carbon beam at a 290 MeV/A bombarding energy and Neon and Argon
beams at 400 MeV/A, hitting Carbon, Copper and Lead ions, are shown
in Fig.~\ref{figneu}, and compared to the experimental data.
Different histograms refer to different emission angles. In particular,
from the top to the bottom of each panel one can distinguish neutrons
emitted at 5$^{\mathrm{o}}$, 10$^{\mathrm{o}}$, 20$^{\mathrm{o}}$, 
30$^{\mathrm{o}}$, 40$^{\mathrm{o}}$, 60$^{\mathrm{o}}$ and 
80$^{\mathrm{o}}$ angles with respect to the incoming beam direction
in the laboratory frame. 
The results at different angles have been multiplied 
by the same factors used in~\citet{iwata} for display purposes.
The results of the theoretical simulations show an overall reasonable agreement
with the experimental data, especially 
at emission angles $\geq 20^{\mathrm{o}}$. On the other side, at forward
emission angles (5$^{\mathrm{o}}$ - 10$^{\mathrm{o}}$), 
the simulations give rise to neutron emission peaks lower than those
experimentally observed. In the case of Carbon projectiles at 290 MeV/A
these peaks are located at higher energies than the experimentally measured
ones, 
and this can be an evidence of the fact that target thickness effects
can not be neglected and have to be included in the simulations. 
In fact, our simulations have been made in the bare approximation
of neglecting all target geometry effects, i.e. each target is 
merely given by a single ion. The same real targets used in the experiments 
with Carbon projectiles at 290 MeV/A have
also been used by~\citet{iwata} with Neon projectiles at 400 MeV/A.
In this last case, 
the agreement of the results of the theoretical simulations 
with the
experimental data is more satisfactory, both regarding the location
of the peaks and their height. 
This point can be ascribed to the fact
that projectiles at higher energies can pe\-ne\-tra\-te more deeply, 
so that ion energy loss and reinteraction effects 
can be neglected more safely.  
To better investigate this aspect, 
we have planned 
to repeat the same simulations, taking into account geometry effects.  
This would be possible thanks to the FLUKA geometry package.
Another source of di\-scre\-pan\-cy could be ascribed to the fact that
isotropic {\it n}-{\it p} and {\it p}-{\it p} 
cross-sections have been used in the simulations,
instead of considering the angular dependence really observed in
nucleon-nucleon scattering experiments. 
We expect that including the last one can further
improve the agreement of the theo\-re\-ti\-cal simulations 
with the experimental data.
As far as intermediate emission angles are concerned, the agreement 
is instead already quite satisfactory. We emphasize that the absolute results
of our simulations shown in Fig.~\ref{figneu} have been obtained 
without making use of any normalization factor. The spectra at intermediate
and backward angles
have been attributed by~\citet{iwata} to the partial
superposition of two components. The equilibrium
component corresponds to the lower kinetic energy ($E_n < $ 20 MeV)
part of the spectra, where an increasing number 
of emitted neutrons  
with decreasing energy
can be observed (shoulder), while the cascade/pre-equilibrium component 
is given by a wide peak extending up to
a few hundreds MeV/A. The theoretical si\-mu\-la\-tions made by QMD + FLUKA 
appear to agree in predicting the presence of both these distributions. 

\subsection{Fragment isotopic distributions}
\label{isot}
\begin{figure}[p!]
\begin{center}
\includegraphics[bb=51 51 550 680,  width=0.99\textwidth]{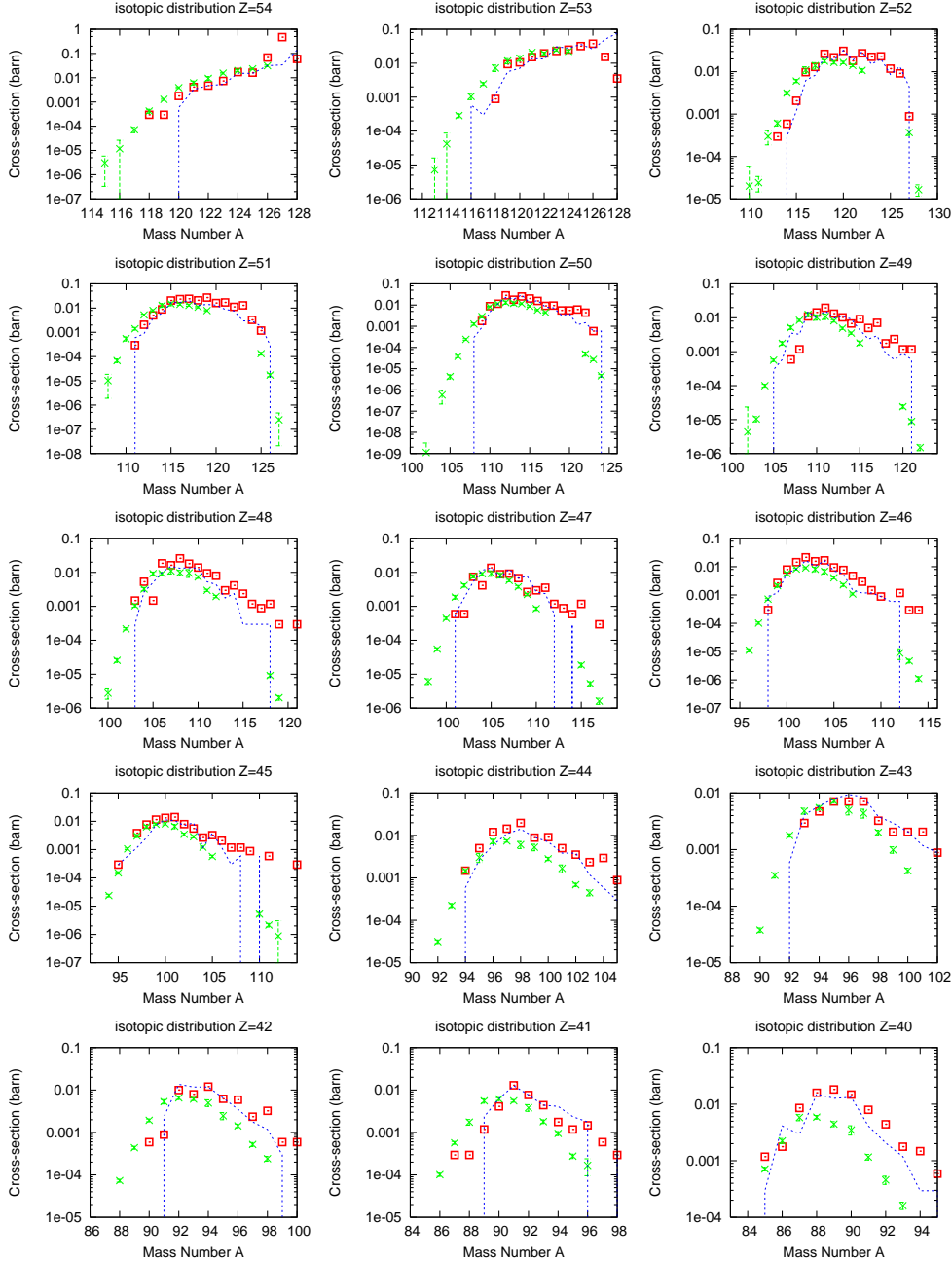}
\caption{\label{figuraiso} Isotopic distributions of projectile-like
fragment production cross-sections (barn) for a 790 MeV/A  
$^{129}\mathrm{Xe}$ beam hitting an Al target. 
Each panel contains the results 
for the isotopes of a different element. The results of the theoretical
simulations of about 10,000 $^{129}\mathrm{Xe}$ + $^{27}\mathrm{Al}$ 
reaction events made by QMD + FLUKA
are shown by open squares, while the experimental
data taken from~\citet{xe790} are shown by crosses.
The results obtained by a modified version of the fully relativistic code
RQMD2.4 + FLUKA are also superimposed on each plot (dotted lines).  
}
\end{center}
\end{figure}

\begin{figure}[p!]
\begin{center}
\includegraphics[bb=51 51 402 300,  width=0.99\textwidth]{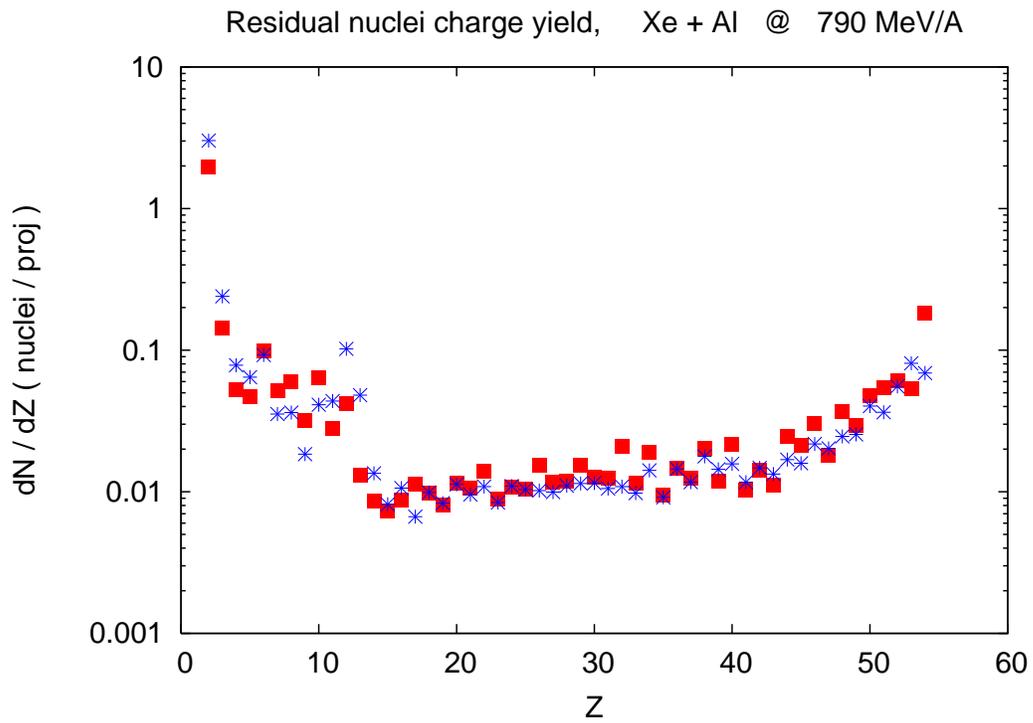}
\caption{\label{spettroz} Charge fragment yield for the same 
projectile-target combination considered in Fig.~\ref{figuraiso}. The
theoretical results predicted by QMD + FLUKA (filled squares) are
compared to those predicted by RQMD2.4 + FLUKA (asterisks).
}
\end{center}
\end{figure}

When two ions collide at energies of a few hundreds MeV/A, highly
excited fragments can be formed, which in turn decay to lower
mass and charge ones by de-excitation processes. Isotopic distributions
for projectile-like fragments 
have been measured by~\citet{xe790} at GSI in the
case of $^{129}\mathrm{Xe}$ ions at a 790 MeV/A bombarding energy, 
impinging on an Aluminium target. We have simulated about 10,000 
$^{129}\mathrm{Xe}$ + $^{27}\mathrm{Al}$ 
reaction events at the same bombarding energy using QMD + FLUKA.
The results of our simulations are presented in Fig.~\ref{figuraiso}.
In each panel the production cross-section is shown for the isotopes
of a different element, ranging from Z = 54 to Z = 40 
(projectile-like isotopes). 
Together with  the results of our theoretical simulations, 
also the experimental measures taken from~\citet{xe790} are plotted.
We can appreciate a better agreement between the theoretical 
and experimental distributions
for the elements whose atomic number is very close to
the projectile Z value (Z~$\sim$~54~-~50), 
than for the lighter
fragments  (Z~$\sim$~44~-~40). 
While the first ones are more copiously produced, the last ones
are more rarely emitted, as follows by comparing their production 
cross-sections. 
The results of our simulations refer to fragment emitted over 
the whole solid angle.
At the considered energy,
higher mass and charge fragments are produced in more peripheral collisions,
and propagate in forward direction, while the lower mass and charge 
ones can be
emitted also at larger angles. 
Furthermore,
at fixed Z, the simulations seem to overestimate the N richer tails, 
while underestimating the higher Z ones. Indeed, an asymmetry between the two
tails is evident also from the experimental data, but to a lesser extent
than obtained by the theoretical simulations.
~\citet{xe790} explain their results suggesting 
that neutron-deficient fragments are formed
from prefragments with high excitation energies, and consequently
a crucial contribution to their formation could be played by evaporation
processes. Since neutrons can evaporate more easily than protons, due
to Coulomb barrier effects on the last ones, 
the evaporation of neutrons would give rise to
these fragments. According to this scenario,
the theoretical excess of N richer fragments could be ascribed
to the fact that the excitation energies of the fragments at the
end of the fast stage of the collision process are not high enough.
We observe that repeating the same simulations by using a completely
different model to describe the fast stage of ion-ion collisions, 
i.e. the relativistic QMD code RQMD2.4 developed
at Frankfurt a few years ago by~\citet{sorge1, sorge2}, 
and already coupled to FLUKA by our 
Collaboration (\citet{aigi}), 
leads to very similar results, also plotted in 
Fig.~\ref{figuraiso}. A comparison between the results of the two
models shows that at a 790 MeV/A bombarding energy
the isotopic distributions are rather insensitive
to the fact that fully relativistic potentials are used, 
or non-relativistic ones, as are the cases of RQMD and QMD, respectively. 
Including a pre-equilibrium phase after the fast stage of the
collisions, with a smooth transition to the
equilibrium stage described by statistical models,
could help in gaining further
insight on the reasons of the discrepancies with experimental data.  
Furthermore, another source of possible discrepancies 
could be the fact
that the Aluminium target thickness has not been taken into account in our
simulations. All these points are currently under investigation. 
The total theoretical charge fragment yield as a function of Z
is presented in Fig.~\ref{spettroz} for completeness, including
also fragments with Z $<$ 40, whose distributions 
have not been measured in the experiment by~\citet{xe790}.

\section{Conclusions}
\label{conclu}
We have performed simulations of ion-ion collision events, in the
approximation of neglecting target geometry effects such as
beam energy loss and re\-in\-te\-rac\-tion during propagation, 
by using the QMD we have developed
in the last few years, interfaced to the statistical de-excitation module
taken from the general purpose Monte Carlo 
FLUKA code. Results of simulations at bombarding energies around
and below 100 MeV/A have been shown is other papers, while here
results have been shown concerning different projectile-target combinations
at energies of a few hundreds MeV/A. Specific observables,
such as neutron double-differential
production cross-sections at different angles and fragment isotopic
distributions, have been calculated. 
Comparisons of the results  with experimental data 
suggest an overall reasonable agreement, even if some discrepancies
also appear, which need further investigation. In particular,
a precise reproduction of the target geometry and the experimental setup,
as well as the inclusion in the model of the angular dependence of
nucleon-nucleon cross-sections, are in our opinion crucial to 
further improve the agreement with the experimental data. 
  




\end{document}